\documentclass[twocolumn,10pt,pra,aps,showpacs]{revtex4-1}
\usepackage{float}
\floatstyle{boxed}
\usepackage{graphicx}
\usepackage{amsbsy}
\usepackage{color}
\usepackage[utf8x]{inputenc}
\usepackage{epstopdf}
\usepackage{amsmath,amssymb,amsfonts}

\begin{document}

\title{Entanglement and nonclassicality in four-mode Gaussian states generated via parametric
down-conversion and frequency up-conversion}
\author{Ievgen I. Arkhipov}
\email{ievgen.arkhipov01@upol.cz}
\address{RCPTM, Joint Laboratory of Optics of Palack\'y University and
Institute of Physics of CAS, Palack\'y University, 17. listopadu
12, 771 46 Olomouc, Czech Republic}

\author{Jan Pe\v{r}ina Jr.}
\address{RCPTM, Joint Laboratory of Optics of Palack\'y University and
Institute of Physics of CAS, Palack\'y University, 17. listopadu
12, 771 46 Olomouc, Czech Republic}

\author{Ond\v{r}ej Haderka}
\address{Institute of Physics of CAS, Joint Laboratory of Optics of Palack\'y University and
Institute of Physic, 17. listopadu 50a, 771 46 Olomouc, Czech
Republic}

\author{Alessia Allevi}
\address{Dipartimento di Scienza e Alta Tecnologia, Universit\`a
degli Studi dell'Insubria, Via Valleggio 11, 22100 Como, Italy}
\address{CNISM UdR Como, Via Valleggio 11, 22100 Como, Italy}

\author{Maria Bondani}
\address{Istituto di Fotonica e Nanotecnologie, Consiglio Nazionale
delle Ricerche, Via Valleggio 11, 22100 Como, Italy}
\address{CNISM UdR Como, Via Valleggio 11, 22100 Como, Italy}

\begin{abstract}
Multipartite entanglement and nonclassicality of four-mode
Gaussian states generated in two simultaneous nonlinear processes
involving parametric down-conversion and frequency up-conversion
are analyzed assuming the vacuum as the initial state. Suitable
conditions for the generation of highly entangled states are
found. Transfer of the entanglement from the down-converted modes
into the up-converted ones is also suggested. The analysis of the
whole set of states reveals that sub-shot-noise intensity
correlations between the equally-populated down-converted modes,
as well as the equally-populated up-converted modes, uniquely
identify entangled states. They represent a powerful entanglement
identifier also in other cases with arbitrarily populated modes.
\end{abstract}

\maketitle

\section*{Introduction}

Since the discovery of quantum mechanics, entanglement has been
considered a very peculiar and purely quantum feature of the
physical systems. Its fundamental importance emerged when the
experiments showing the violation of the Bell
inequalities~\cite{Aspect82,Weihs1998,Brunner14}, implementing
quantum teleportation~\cite{Bouwmeester1997,Genovese2005} or
demonstrating dense coding were performed. Nowadays, entanglement
is undoubtedly considered as the key resource of modern and
emerging quantum technology, including quantum metrology, quantum
computation~\cite{NielsenBook} and quantum
communications~\cite{aoki03,loock01,yonezawa04}.

For this reason, a great deal of attention has been devoted to the
construction of practical sources of entangled light, both in the
domains of discrete and continuous variables. While individual
entangled photon pairs arising in spontaneous parametric
down-conversion are commonly used in the discrete domain
\cite{Bouwmeester2000}, single-mode as well as two-mode squeezed
states originating in parametric down-conversion and containing
many photon pairs represent the sources in the domain of
continuous variables \cite{Dodonov2002}. Even more complex
nonlinear optical processes, including those combining
simultaneous parametric down-conversion and frequency
up-conversion, have been analyzed as sources of more complex
entangled states. This approach has been experimentally
implemented in Refs.~\cite{alessia04,coelho} considering
three-mode entanglement and in Ref.~\cite{pysher} where the
four-mode entanglement has been analyzed.

Here, we consider a four-mode system composed of two
down-converted modes and two up-converted modes. In the
system, parametric down-conversion and frequency up-conversion
involving both down-converted modes simultaneously occur in the
same nonlinear medium \cite{Boyd2003}. While parametric
down-conversion serves as the primary source of entanglement
\cite{Mandel1995}, frequency up-conversion is responsible for the
transfer of the entanglement to the up-converted modes.

This transfer operation is interesting from the fundamental point
of view, as it generalizes the well-known property of `one-mode'
frequency up-conversion pumped by a strong coherent field, in
which the statistical properties of the incident field are
transferred to the frequency up-converted counterpart, also
including the nonclassical ones (e.g., squeezing,
\cite{Perina1991}). We note that such properties are important for
the applications of the up-conversion process: For instance, it
has been used many times for `shifting' an optical `one-mode'
field to an appropriate frequency where its detection could be
easily achieved~\cite{Langrock2005,Zeilinger12}.

In the general analysis of the four-mode system, we quantify its
global nonclassicality via the Lee nonclassicality depth
\cite{Lee91}. Since the four-mode system under consideration
cannot exhibit nonclassicality of individual single modes, the
global nonclassicality automatically implies the presence of
entanglement among the modes (for a two-mode Gaussian system
involving parametric down-conversion, see \cite{Arkhipov2016}).
The analysis of `the structure of entanglement' further simplifies
by applying the Van Loock and Furusawa inseparability
criterion~\cite{vanLoock03} that excludes the presence of genuine
three- and four-partite entangled states. This means that in the
system discussed here there are only bipartite entangled states.
It is thus sufficient to divide the analyzed four-mode state into
different bipartitions to monitor the structure of entanglement.
Then, the well-known entanglement criterion based on the positive
partial transposition of the statistical
operator~\cite{Peres96,Horodecki97}, which gives the logarithmic
negativity as an entanglement quantifier, is straightforwardly
applied~\cite{Hill1997,Vidal02}.

The experimental detection of two-mode (-partite) entanglement is
in general quite challenging, as it requires measurements in
complementary bases. Here, we theoretically show that, for the
considered system with the assumed initial vacuum state, any
two-mode partition exhibiting sub-shot-noise intensity
correlations is also entangled. As a consequence, the measurement
of intensity auto- and cross-correlations in this system is
sufficient to give the evidence of the presence of two-mode
entangled states through the commonly used noise reduction factor.

Finally, we note that the Hamiltonian of the analyzed four-mode
system formally resembles that describing a twin beam with signal
and idler fields divided at two beam splitters. This analogy
results in similar properties of the four-mode states obtained in
the two cases, though the processes of down-conversion and
up-conversion occur simultaneously in our system, at variance with
the system with two beam splitters, which modify the already
emitted twin beam. We note that the system with two beam splitters
has been frequently addressed in the literature as a prototype of
more complex devices based on two multiports that are used to have
access to intensity correlation functions for the detailed
characterization of the measured fields \cite{Vogel2008}, also
including their photon-number statistics
\cite{Waks2004,Haderka2005a,Avenhaus2008,PerinaJr2012,Sperling2012,Allevi2012}.

The paper is organized as follows. In Section {\it Four-mode
nonlinear interaction} the model of four-mode nonlinear
interaction including parametric down-conversion and frequency
up-conversion is analyzed. Nonclassicality of the overall system
is addressed in Section {\it Nonclassicality}. In Section {\it
Four-mode entanglement}, the entanglement of the overall system is
investigated considering the partitioning of the state into
different bipartitions. Two-mode entangled states obtained after
state reduction are analyzed in Section {\it Two-mode entanglement
and noise reduction factor}, together with two-mode sub-shot-noise
intensity correlations. Suitable parameters of the corresponding
experimental setup can be found in Section {\it Experimental
implementation}. Section {\it Conclusions} summarizes the obtained
results.

\section*{Four-mode nonlinear interaction}

We consider a system of four nonlinearly interacting optical modes
(for the scheme, see Fig.~1). Photons in modes 1 and 2 are
generated by parametric down-conversion with strong pumping
(coupling constant $ g_1 $). Photons in mode 1 (2) can then be
annihilated with the simultaneous creation of photons in mode 3
(4). The two up-conversion processes are possible thanks to the
presence of two additional strong pump fields with coupling
constants $ g_2 $ and $ g_3 $. The overall interaction Hamiltonian
for the considered four-mode system is written as \cite{Boyd2003}:
\begin{equation}\label{hamil}
 {\hat H}_{\mathrm{int}}=\hbar g_{1}{\hat a_{1}}^{\dagger}{\hat a_{2}}^{\dagger}+
  \hbar g_{2}\hat a_{1}{\hat a_{3}}^{\dagger}+\hbar g_{3}\hat a_{2}{\hat a_{4}}^{\dagger}+\mathrm{H.c.},
\end{equation}
where the operators $\hat a_{1}^\dagger$ and $\hat a_{2}^\dagger $
create an entangled photon pair in modes 1 and 2 and the creation
operators $\hat a_{3}\dagger$ and $\hat a_{4}^\dagger $ put the
up-converted photons into modes 3 and 4, respectively. Symbol $
{\rm H.c.} $ replaces the Hermitian conjugated terms.
\begin{figure}  
\includegraphics[width=0.27\textwidth]{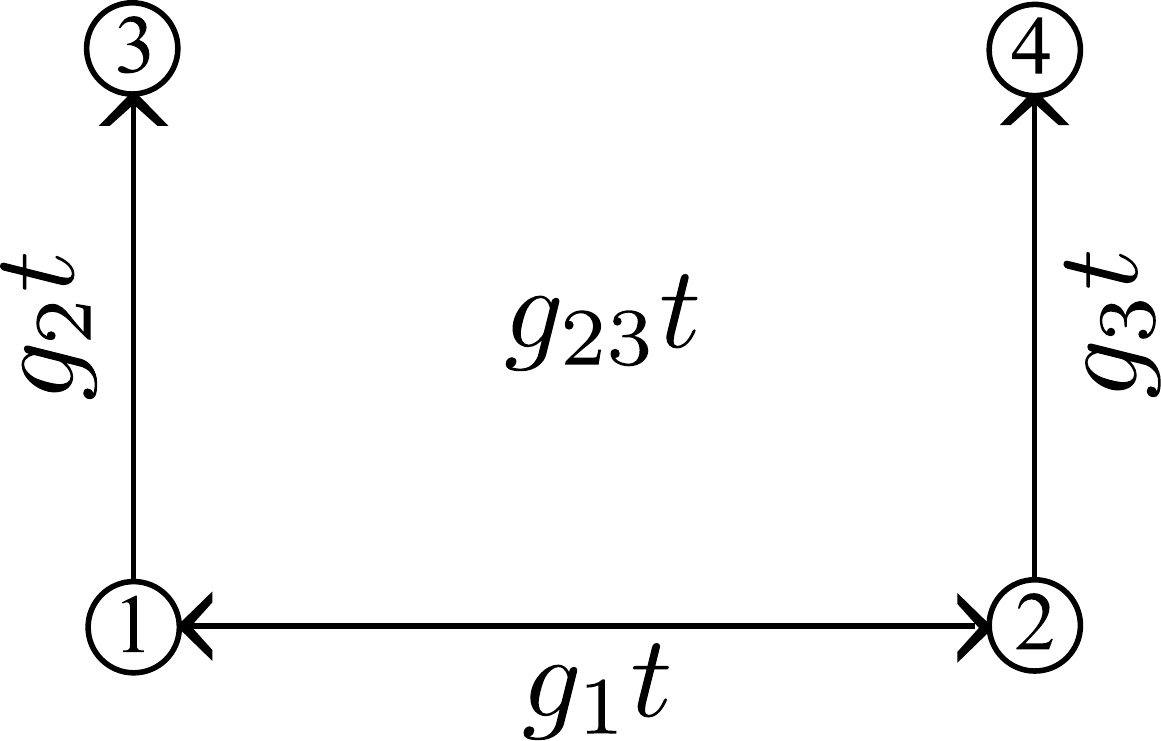}
 \caption{Optical fields in modes 1 and 2 interact via parametric down-conversion
 described by the nonlinear coupling constant $ g_1 $. Photons from mode 1 (2) are
 converted into photons of mode 3 (4) thanks to the frequency
 up-conversion characterized by the coupling constant $ g_2 $ ($ g_3
 $); $ t $ stands for the interaction time. In the symmetric case we have $ g_{23} = g_2 = g_3
 $.}
\end{figure}

The Heisenberg-Langevin equations corresponding to the Hamiltonian
$ \hat H_{\mathrm{int}} $ in Eq.~(\ref{hamil}) are written in
their matrix form as follows:
\begin{eqnarray} \label{hle}    
 \frac{d \hat {\bf a}}{dt} &=& {\bf U} \hat {\bf a} + \hat {
 \bf L},
\end{eqnarray}
where $\hat {\bf a} = (\hat a_1^{\dagger},\hat a_2,\hat
a_3^{\dagger},\hat a_4 )^T$ and $\hat {\bf L} = (\hat
L_1^{\dagger},\hat L_2,\hat L_3^{\dagger},\hat L_2 )^T$. The
matrix $ {\bf U} $ introduced in Eq.~(\ref{hle}) is expressed as
\begin{equation}   
 {\bf U} = \begin{pmatrix}
 -\gamma_1/2 & -ig_1 & -ig_2 & 0 \\
  ig_1 &  -\gamma_2/2 & 0 & ig_3 \\
   -ig_2 & 0 &  -\gamma_3/2& 0 \\
    0 & ig_3 & 0 & -\gamma_4/2,
 \end{pmatrix}
 \end{equation}
in which $\gamma_j$ stands for the damping coefficient of mode $ j
$, $ j=1,\ldots,4 $. The Langevin operators $\hat L_j $, $
j=1,\ldots,4 $, obey the following relations:
\begin{eqnarray}            
 \langle\hat L_{j}(t)\rangle = \langle\hat
  L^{\dagger}_{j}(t)\rangle= 0, \quad
 \langle\hat L^{\dagger}_{j}(t)\hat L_{k}(t')\rangle =
  \delta_{jk}\gamma_j \langle n_{dj}\rangle\delta(t-t'), \quad
 \langle\hat L_{j}(t)\hat L^{\dagger}_{k}(t')\rangle =\delta_{jk}\gamma_j ( \langle n_{dj}\rangle
  +1)\delta(t-t').
\label{4}
\end{eqnarray}
The Kronecker symbol is denoted as $\delta_{ij}$ and the symbol
$\delta(t)$ means the Dirac function. The mean numbers $ n_{dj} $
corresponding to noise reservoir photons have been used in
Eqs.~(\ref{4}). We note that for the noiseless system the
following quantity $ \langle \hat{a}_1^\dagger \hat{a}_1 \rangle +
\langle \hat{a}_4^\dagger \hat{a}_4 \rangle - \langle
\hat{a}_2^\dagger \hat{a}_2 \rangle - \langle \hat{a}_3^\dagger
\hat{a}_3 \rangle $ is conserved in the interaction.

Introducing frequencies $ \omega_j $ and wave vectors $\vec{k}_j $
of the mutually interacting modes, we formulate the assumed ideal
frequency and phase-matching conditions of the considered
nonlinear interactions in the form:
\begin{eqnarray}\label{5}  
 && \omega_{p12}=\omega_1+\omega_2, \quad \omega_{p13}=\omega_1+\omega_3, \quad \omega_{p24}=\omega_2+\omega_4, \nonumber \\
 && \vec{k}_{p12}=\vec{k}_1+\vec{k}_2, \quad \vec{k}_{p13}=\vec{k}_1+\vec{k}_3, \quad  \vec{k}_{p24}=\vec{k}_2+\vec{k}_4.
\end{eqnarray}
In Eqs.~(\ref{5}), $ \omega_{p12}$ ($\vec{k}_{p12}$) stands for
the pump-field frequency (wave vector) of parametric
down-conversion, whereas $ \omega_{p13}$ [$ \omega_{p24}$]
($\vec{k}_{p13}$ [$\vec{k}_{p24}$]) means the frequency (wave
vector) of the field pumping the up-conversion process between
modes 1 [2] and 3 [4].

The solution of the system of first-order linear operator
stochastic equations~(\ref{hle}) can be conveniently expressed in
the following matrix form:
\begin{equation}\label{A}
\hat {\bf a}(t) = {\bf M} \hat {\bf a}(0) + \hat {\bf F}(t),
\end{equation}
where the evolution matrix $ \bf M $ is written in Eq.~(\ref{M})
of Appendix for the noiseless case and vector $\hat F$ arises from
the presence of the stochastic Langevin forces. More details can
be found in Ref.~\cite{PerinaJr2000}. When applying the solution
(\ref{A}), we consider the appropriate phases of the three pump
fields such that the coupling constants $ g_j $, $ j=1,2,3 $, are
real.

The statistical properties of the optical fields generated both by
parametric down-conversion and up-conversion are described by the
normal characteristic function $ C_{\mathcal N} $ defined as
\begin{eqnarray}\label{ncf}    
C_{\mathcal N}(\mbox{\boldmath$ \beta $}) =
\mathrm{Tr}\left[\hat\rho(0)\exp \left(\sum_{i=1}^{4}\beta_{i}\hat
a^{\dagger}_{i}\right)\exp\left(-\sum_{i=1}^4\beta^{\ast}_{i}\hat
a_{i}\right)\right], \nonumber \\
\end{eqnarray}
where ${\rm Tr}$ denotes the trace and $ \mbox{\boldmath$ \beta $}
\equiv (\beta_1,\beta_2,\beta_3,\beta_4)^T $. Using the solution
given in Eq.~(\ref{A}), the normal characteristic function $
C_{\mathcal N} $ attains the Gaussian form:
\begin{eqnarray} \label{acf}      
 C_{\mathcal N}(\mbox{\boldmath$ \beta $})&=&\exp\Big\{
  -\sum_{i=1}^{4}B_{i}\vert\beta_{i}\vert^{2}
  + \Bigl[ D^{\ast}_{12}\beta_{1}\beta_{2}+\bar D^{\ast}_{13}\beta_{1}\beta_{3}^{\ast}+ \nonumber  \\
 && D^{\ast}_{14}\beta_{1}\beta_{4}
  +D^{\ast}_{23}\beta_{2}\beta_{3}+\bar
  D^{\ast}_{24}\beta_{2}\beta_{4}^{\ast}+\mathrm{c.c.} \Bigr] \Big\}
\end{eqnarray}
and $ {\rm c.c.} $ replaces the complex conjugated terms. The
coefficients occurring in Eq.~(\ref{acf}) are derived in the form:
\begin{eqnarray}\label{param}
B_{1}&=&\langle\Delta\hat a_1^{\dagger}\Delta\hat a_1\rangle=\vert M_{12}\vert^2+\vert M_{14}\vert^2+\langle\hat F_1^{\dagger}\hat F_1\rangle, \nonumber \\
B_{2}&=&\langle\Delta\hat a_2^{\dagger}\Delta\hat a_2\rangle=\vert M_{21}\vert^2+\vert M_{23}\vert^2+\langle\hat F_2^{\dagger}\hat F_2\rangle, \nonumber \\
B_{3}&=&\langle\Delta\hat a_3^{\dagger}\Delta\hat a_3\rangle=\vert M_{32}\vert^2+\vert M_{34}\vert^2+\langle\hat F_3^{\dagger}\hat F_3\rangle, \nonumber \\
B_{4}&=&\langle\Delta\hat a_4^{\dagger}\Delta\hat a_4\rangle=\vert M_{41}\vert^2+\vert M_{43}\vert^2+\langle\hat F_4^{\dagger}\hat F_4\rangle, \nonumber \\
 D_{12}&=&\langle\Delta\hat a_1\Delta\hat a_2\rangle=M_{11}^{\ast}M_{21}+M_{13}^{\ast}M_{23}+\langle\hat F_1\hat F_2\rangle, \nonumber \\
\bar  D_{13}&=&-\langle\Delta\hat a_1^{\dagger}\Delta\hat a_3\rangle=-M_{11}^{\ast}M_{31}-M_{13}^{\ast}M_{33}-\langle\hat F_1^{\dagger}\hat F_3\rangle, \nonumber \\
 D_{14}&=&\langle\Delta\hat a_1\Delta\hat a_4\rangle=M_{11}^{\ast}M_{41}+M_{13}^{\ast}M_{43}+\langle\hat F_1\hat F_4\rangle, \nonumber \\
 D_{23}&=&\langle\Delta\hat a_2\Delta\hat a_3\rangle=M_{32}^{\ast}M_{22}+M_{34}^{\ast}M_{24}+\langle\hat F_2\hat F_3\rangle, \nonumber \\
\bar  D_{24}&=&-\langle\Delta\hat a_2^{\dagger}\Delta\hat a_4\rangle=-M_{42}^{\ast}M_{22}-M_{44}^{\ast}M_{24}-\langle\hat F_2^{\dagger}\hat F_4\rangle, \nonumber \\
 D_{34}&=&\langle\Delta\hat a_3\Delta\hat a_4\rangle=M_{31}^{\ast}M_{41}+M_{33}^{\ast}M_{43}+\langle\hat F_1\hat F_4\rangle. \nonumber \\
\end{eqnarray}
We note that the two-mode interactions characterized by the
coefficients $ D_{ij} $ and $ \bar D_{ij} $ in Eq.~(\ref{acf})
attain specific forms. While the coefficients $ D_{ij} $ reflect
the presence of photon pairs in modes $ i $ and $ j $,
coefficients $ \bar D_{ij} $ describe mutual transfer of
individual photons between modes $ i $ and $ j $.

The normal characteristic function $C_{\mathcal N}$ can be
rewritten in the matrix form $
\exp(\mbox{\boldmath$\beta$}^\dagger {\bf A}
\mbox{\boldmath$\beta$} /2)$ by introducing the normally-ordered
covariance matrix $ {\bf A} $:
\begin{equation}\label{CM}      
 {\bf A} =
\begin{pmatrix}
 {\bf A}_{1} & {\bf D}_{12} & {\bf D}_{13} & {\bf D}_{14} \\
 {\bf D}_{12}^\dagger & {\bf A}_{2} & {\bf D}_{23} & {\bf D}_{24} \\
 {\bf D}_{13}^\dagger & {\bf D}_{23}^\dagger  & {\bf A}_{3} & {\bf D}_{34} \\
 {\bf D}_{14}^\dagger & {\bf D}_{24}^\dagger  & {\bf D}_{34}^\dagger & {\bf A}_{4}
\end{pmatrix},
\end{equation}
where the $2\times2$ matrices are defined as:
\begin{eqnarray}
 {\bf A}_i &=& \begin{pmatrix}
   -B_i & 0 \\
   0 & -B_i
   \end{pmatrix}, \quad i=1,\ldots,4, \nonumber \\
 {\bf D}_{jk} &=& \begin{pmatrix}
   \bar D_{jk}^{*} &  D_{jk} \\
    D_{jk}^{*} & \bar D_{jk}
   \end{pmatrix}, \quad j,k=1,\ldots,4.
\end{eqnarray}

The covariance matrix $ \mbox{\boldmath$\sigma$} $ related to the
symmetric ordering and corresponding to the phase space $(\hat x,
\hat p)$ is needed to perform easily partial transposition. It has
the same structure as the covariance matrix $ {\bf A} $ written in
Eq.~(\ref{CM}) with the blocks $ {\bf A}_i $ ($ {\bf D}_{jk} $)
replaced by the blocks $ \mbox{\boldmath$\sigma_i$} $ ($
\mbox{\boldmath$\varepsilon_{jk}$} $) defined as:
\begin{eqnarray}
 \mbox{\boldmath$\sigma_i$}&=&\begin{pmatrix}
  B_i+\frac{1}{2} & 0 \\
  0 & B_i+\frac{1}{2}
  \end{pmatrix},  \nonumber \\
 \mbox{\boldmath$\varepsilon_{jk}$} &=&\begin{pmatrix}
  {\mathrm {Re}}(D_{jk}-\bar
   D_{jk}) &  {\mathrm {Im}}(D_{jk}-\bar D_{jk}) \cr
  {\mathrm {Im}}(D_{jk}+\bar D_{jk}) & - {\mathrm{Re}}(D_{jk}+\bar
  D_{jk})
 \end{pmatrix}, \nonumber \\
 & &  \hspace{3cm} i,j,k=1,\ldots,4.
\end{eqnarray}
Symbol $ {\rm Re} $ ($ {\rm Im} $) denotes the real (imaginary)
part of the argument.

In what follows, we consider the situation in which all four modes
begin their interaction in the vacuum state. Moreover, we focus on
the specific symmetric case in which $ g_2 = g_3 \equiv g_{23} $.
A note concerning the general case $ g_2 \neq g_3 $ is found at
the end.

\section*{Nonclassicality}

We first analyze the global nonclassicality of the whole four-mode
system as it is relatively easy and, for the considered initial
vacuum state, it implies entanglement (see below). Nonclassicality
of the whole four-mode state described by the statistical operator
$\hat\rho$ is conveniently quantified by the Lee nonclassicality
depth $\tau$ \cite{Lee91}. This quantity gives the amount of
noise, expressed in photon numbers, needed to conceal nonclassical
properties exhibited by the Glauber-Sudarshan $P$ function, which
attains negative values in certain regions or even does not exist
as an ordinary function. The Glauber-Sudarshan $P$ function is
determined by the Fourier transform of the normally-ordered
characteristic function $ C_{\cal N} $ given in Eq.~(\ref{acf}).
Technically, the Lee nonclassicality depth is given by the largest
positive eigenvalue of the covariance matrix $ {\bf A} $ defined
in Eq.~(\ref{CM}). So, it can be easily determined.

The Lee nonclassicality depth $ \tau $ as a function of the
coupling parameters $g_1t$ and $g_{23}t$ is shown in Fig.~2. 
\begin{figure} 
 \includegraphics[width=0.4\textwidth]{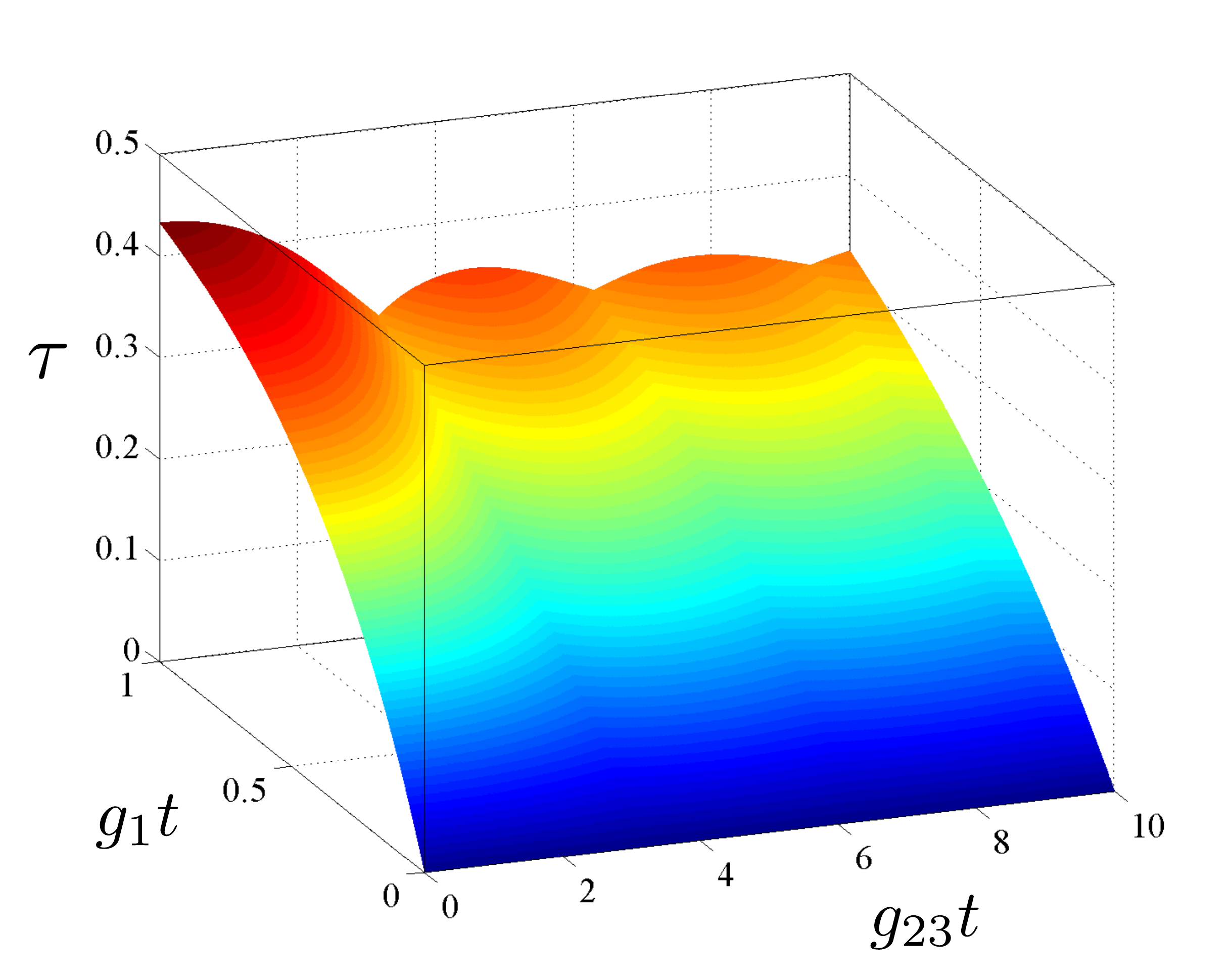}
 \caption{ Nonclassicality depth $\tau$ as { a function of the} parameters
 $g_1t$ and $g_{23}t$.}
\end{figure}
The
increasing values of $g_1t$ result in larger values of the
nonclassicality depth $\tau$, as the number of photons
simultaneously generated in modes 1 and 2 increases. We note that
this pairing of photons in the process of parametric
down-conversion is the only source of nonclassicality in the
analyzed four-mode system. On the contrary, nonzero values of
parameter $g_{23}t$ only lead to the oscillations of the
nonclassicality depth $ \tau $. This behavior occurs as the
frequency up-conversion moves photons, and so also photon pairs,
from modes 1 and 2 to modes 3 and 4 and vice versa (see the scheme
in Fig.~1). This results in the nonclassical properties of modes 3
and 4, at the expenses of the nonclassical properties of modes 1
and 2.

The maximum value of the Lee nonclassicality depth $ \tau = 0.5 $
is reached for $ g_{23}t = 0 $ and ideally in the limit $ g_1 t
\rightarrow \infty $, i.e. when only the strong parametric
down-conversion occurs. This is in agreement with the analysis of
nonclassical properties of twin beams reported
in Ref.~\cite{arkhipov15}. The value $ \tau=0.5 $ can also be
asymptotically reached in the limit $ g_{23}t \rightarrow \infty
$, in which we have
\begin{equation}\label{tau_c}
 \tau_{g_{23}t\to\infty} = \frac{1}{2}\left[\sqrt{(B_{1}-B_{2})^2+4\vert D_{12}\vert^2} -
  (B_{1}+B_{2})\right]
\end{equation}
with $ B_3 \rightarrow B_1 $, $ B_4 \rightarrow B_2 $ and $ D_{34}
\rightarrow D_{12} $. It is worth noting that
formula~(\ref{tau_c}) applies also for $ g_{23}t = 0 $.

Nonclassicality is also strongly resistant against damping in the
system. This means that even a low number of photon pairs is
sufficient to have a nonclassical state. We demonstrate this
resistance by considering the damping constants $ \gamma $
proportional to the nonlinear coupling constant $ g_1 $, which
quantifies the speed of photon-pair generation. The graphs in
Fig.~3 show that the generated states remain strongly nonclassical
even though a considerable fraction of photon pairs is broken
under these conditions. The comparison of graphs in Figs.~3(a) and
(b) reveals that the damping is more detrimental in the
down-converted modes 1 and 2 than in the up-converted modes 3 and
4.

At variance with nonclassicality, the determination and
quantification of entanglement is more complex and it is
technically accomplished by considering all possible bipartitions
of the four-mode system (see the next Section). On the one side
all bipartitions considered below are in principle sufficient to
indicate entanglement, on the other side the application of the
Van Loock and Furusawa inseparability criterion \cite{vanLoock03}
to our system excludes the presence of genuine three- and
four-mode entanglement. The analyzed Hamiltonian written in
Eq.~(\ref{hamil}) together with the incident vacuum state also
excludes the presence of nonclassical states in individual modes.
In what follows, the bipartite entanglement is thus the only
source of the global nonclassicality in the analyzed system. This
situation considerably simplifies the possible experimental
investigations as positive values of the Lee nonclassicality depth
directly imply the presence of entanglement somewhere in the
system.
\begin{figure} 
\includegraphics[width=0.5\textwidth]{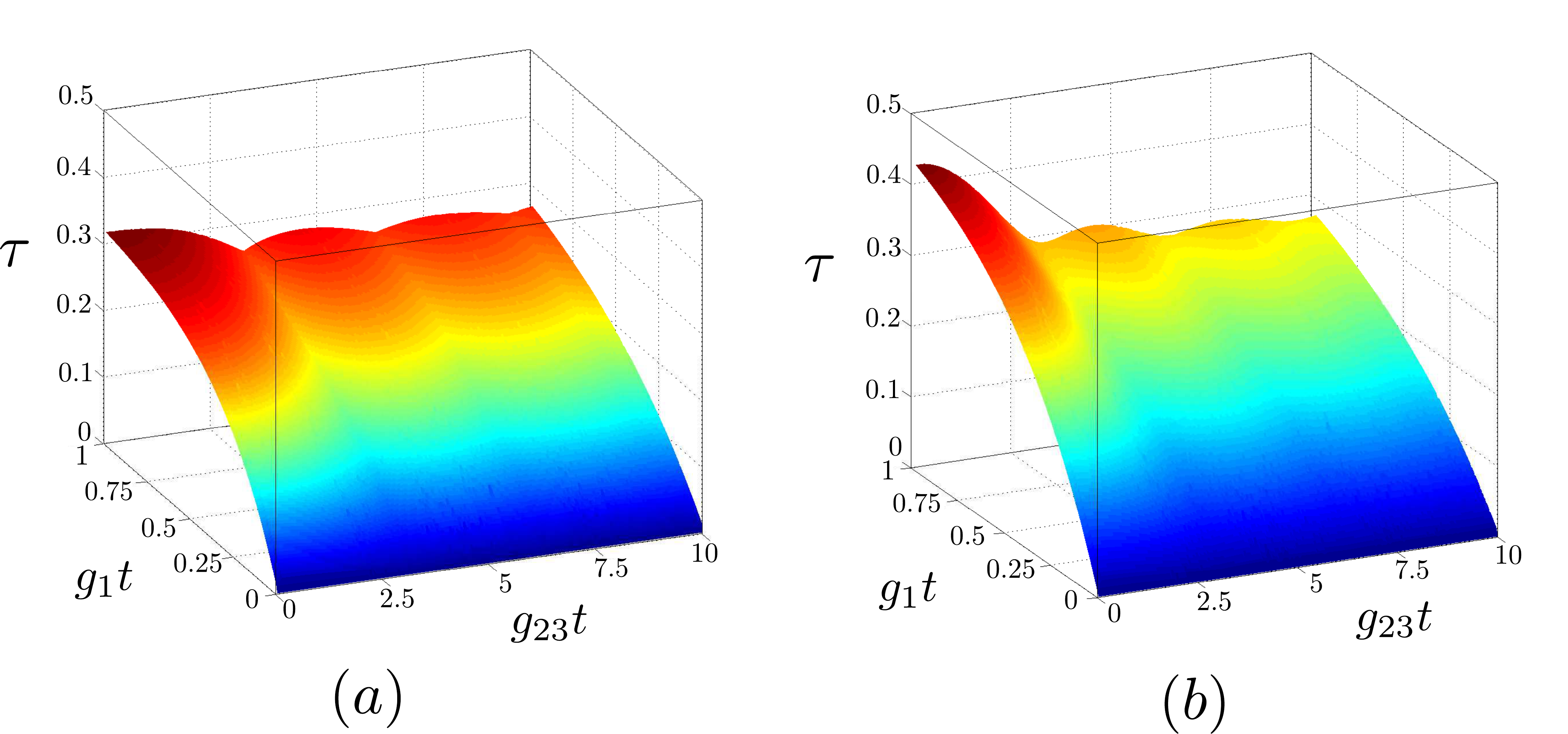}
 \caption{Nonclassicality depth $\tau$ as it depends on parameters
 $g_1t$ and $g_{23}t$ for (a) $\gamma_1t=\gamma_2t= g_1 t $, $\gamma_3t=\gamma_4t=0$; (b)
 $\gamma_1t=\gamma_2t=0$, $\gamma_3t=\gamma_4t=g_1t$, assuming
  $n_{dj}=0$ for $ j=1,\ldots,4 $.}
\label{damp}
\end{figure}

\section*{Four-mode entanglement}

In quantifying the entanglement in our four-mode Gaussian system,
we rely on the following facts applicable to an arbitrary
$(m+n)$-mode Gaussian state. It has been proven that positivity of
the partially transposed (PPT) statistical operator describing any
$ 2 \times 2 $ or $ 2 \times 3 $ bipartition of the state is a
necessary condition for the separability of the
state~\cite{Peres96,Horodecki97}. Moreover, it has been shown that
the violation of PPT condition occurring in any $1\times(m+n-1)$
bipartitions or $m\times n$ bisymmetric bipartitions for $ m
>2 $ and $ n > 3 $ is a sufficient condition for the
entanglement in the analyzed $(m+n)$-mode
state~\cite{Simon00,Serafini05}. For continuous variables systems,
the PPT is simply accomplished when the symmetrically-ordered
field operators are considered allowing to perform the PPT only by
changing the signs of the momenta $\hat p$~\cite{Simon00}.
Moreover, symplectic eiganvalues $\tilde n_i$ of the
symmetrically-ordered covariance matrix $ \mbox{\boldmath$ \sigma
$} $ can be conveniently used to quantify entanglement in
bipartite systems via the logarithmic negativity $ E
$~\cite{Vidal02}, defined in terms of eigenvalues $\tilde
n_i<1/2$:
\begin{equation}\label{LN}
 E  = \max\left\{0,-\sum\limits_{i} {\mathrm {log}}(2\tilde n_i)\right\},
\end{equation}
where $ {\rm  max} $ gives the maximal value.

\begin{figure}   
\includegraphics[width=0.5\textwidth]{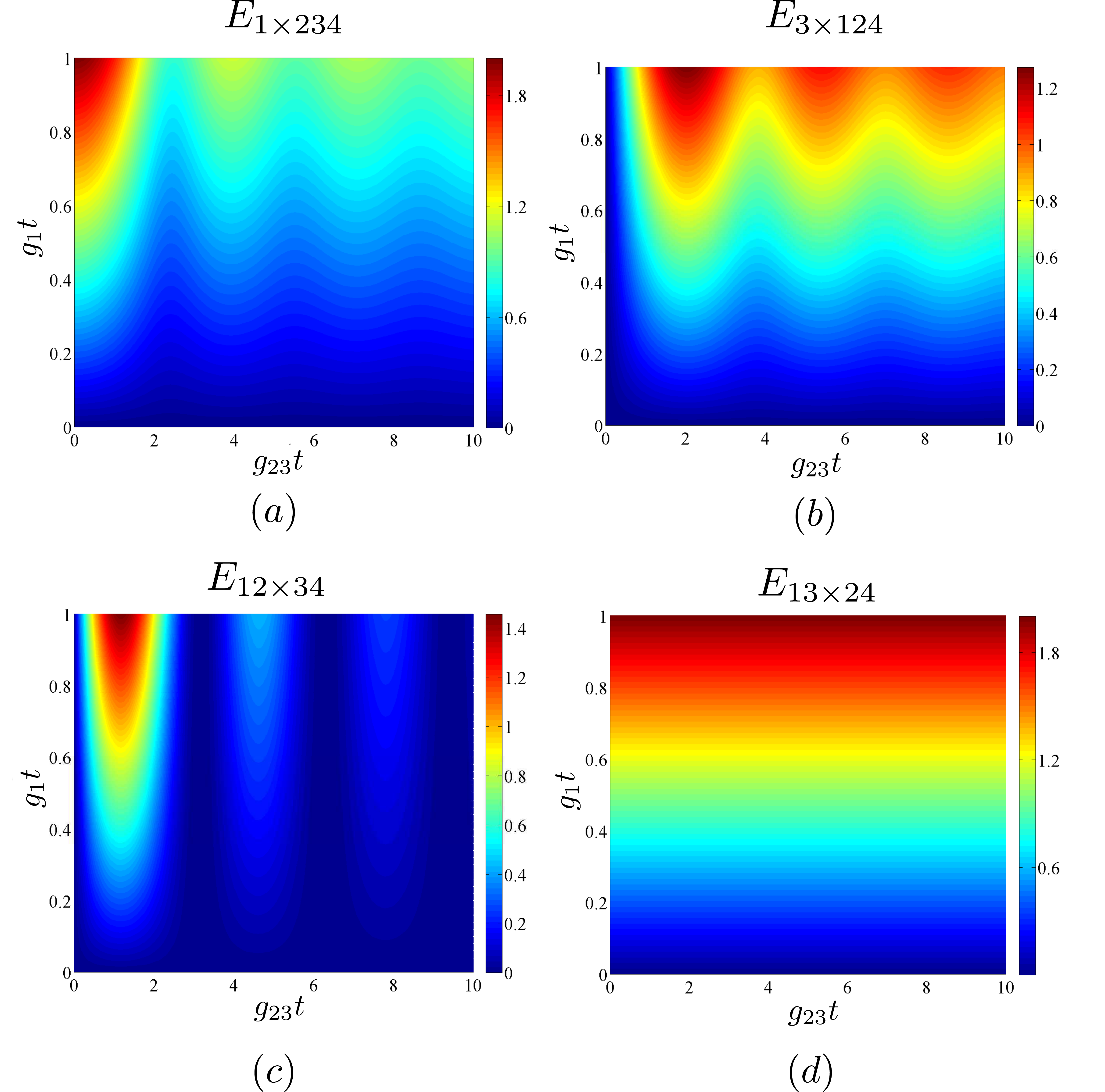}
 \caption{Logarithmic negativities $E_{1\times234}$ (a),
  $E_{3\times124}$ (b), $E_{12\times34}$ (c), and $E_{13\times24}$
  (d) as functions of parameters $g_1t$ and $g_{23}t$ for different
  bipartitions indicated in the subscripts.}
\end{figure}

In the four-mode Gaussian state sketched in Fig.~1, we have two
kinds of bipartitions. Either a single mode forms one subsystem
and the remaining three modes belong to the other subsystem, or
two modes are in one subsystem and the remaining two modes lie in
the other subsystem. Due to the symmetry, only two members of each
group are of interest for us. Namely, these are bipartitions $
1\times 234 $ and $ 3\times 124 $ from the first group and
bipartitions $ 12\times 34 $ and $ 13\times 24 $ from the second
one. We note that, while the bipartition $ 12\times 34 $ is
bisymmetric in our interaction configuration (provided that $ g_2
t=g_3 t $), the bipartition $ 13\times 24 $ is not bisymmetric.
Nevertheless, positive values of both the logarithmic negativities
$ E_{12\times34} $ and $ E_{13\times24} $ reflect entanglement as
both bipartitions involve two modes on both sides. Similarly,
positive values of the logarithmic negativities $ E_{1\times234} $
and $ E_{3\times124} $ guarantee the presence of entanglement.

\begin{figure} 
\includegraphics[width=0.45\textwidth]{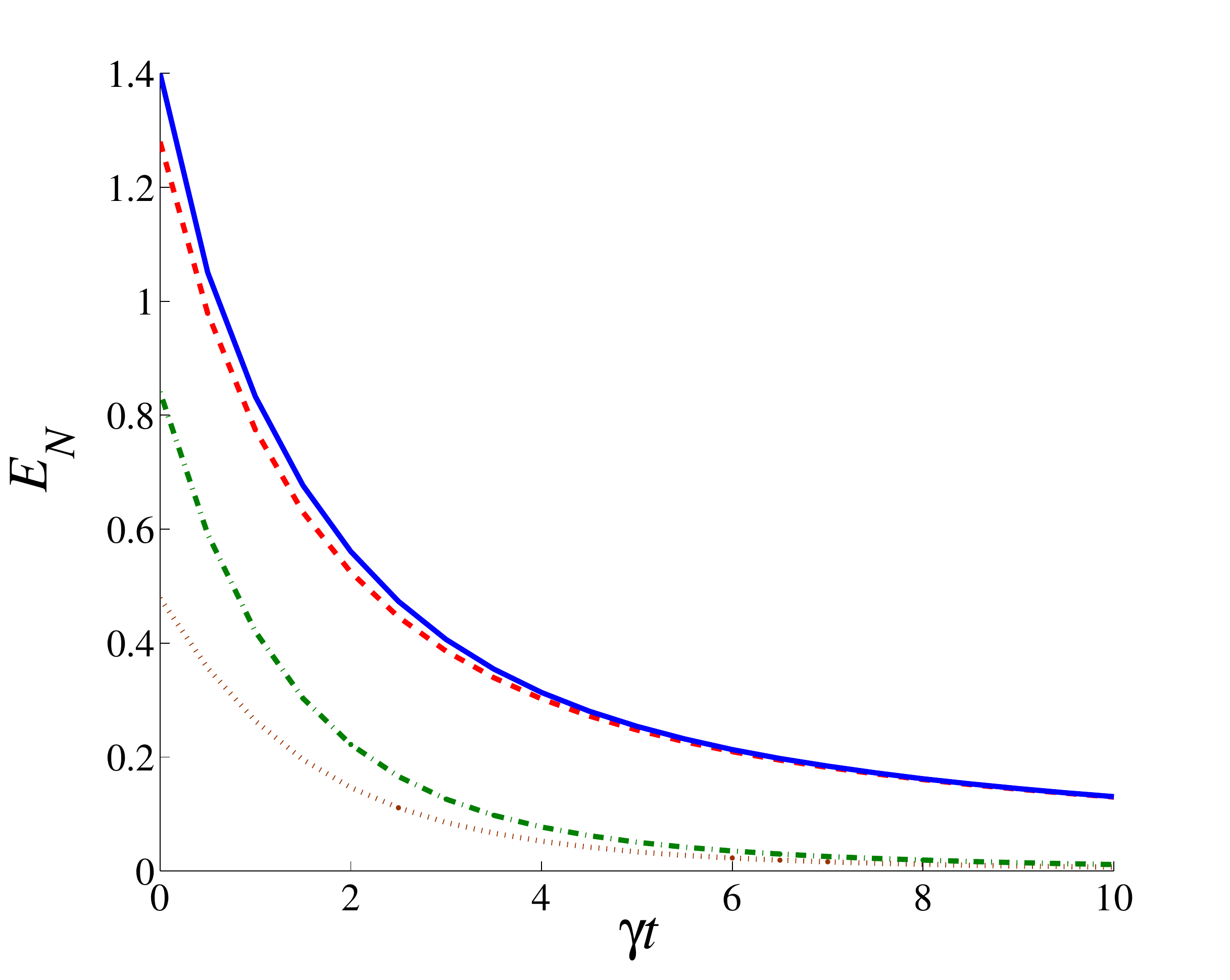}
\caption{Logarithmic negativity $E$ as a function of the damping
coefficient $\gamma t$ for different bipartitions: $1\times234$
(dashed red line), $3\times124$ (brown dotted line), $12\times34$
(dashed-dotted green line), and $13\times24$ (solid blue line). We
set $g_1t=g_2t=g_3t=0.7$, $ \gamma \equiv \gamma_1 = \gamma_2 =
\gamma_3 = \gamma_4 $; $ n_{dj}=0 $ for $ j=1,\ldots,4 $.}
\end{figure}

We first pay attention to the entanglement expressed in the
logarithmic negativities $ E_{1\times234} $ and $ E_{3\times124}
$. As suggested by the graphs in Figs.~4(a) and (b), the
oscillating behavior of negativity $E_{1\times234} $ is
complementary to that of negativity $ E_{3\times124} $. This means
that the larger values of negativity $E_{1\times234}$ are
accompanied by the lower values of negativity $ E_{3\times124}$
and vice versa. Such a result is a consequence of the fact that
the entanglement is due to the presence of photon pairs and a
photon created in mode 1 can move to mode 3 and later return back
to mode 1. This movement leads to the oscillations with frequency
$ g_{23} $, which are clearly visible in Figs.~4(a) and (b). This
explanation also suggests that no entanglement is possible between
modes 1 and 3. Indeed, if we also determine the negativity $
E_{1\times24} $ (or $ E_{3\times24} $), we will get the same
values already obtained for the negativity $ E_{1\times234} $ ($
E_{3\times124} $).

The negativity $ E_{12\times34} $, characterizing the entanglement
between the twin beam in modes 1 and 2 and the up-converted beams
in modes 3 and 4, is plotted in Fig.~4(c). It reflects the gradual
movement of photon pairs from modes 1 and 2, where they are
created, to modes 3 and 4. Note that the maxima of negativity $
E_{12\times34} $ along the $ g_{23}t $-axis occur inbetween the
maxima of negativities $ E_{1\times234} $ and $ E_{3\times124} $.
The origin of entanglement in photon pairing is confirmed in the
graph of Fig.~4(d), showing that the negativity $ E_{13\times24} $
is independent of parameter $ g_{23}t $ and that the negativity $
E_{13\times24} $ increases with the increasing parameter $ g_1 t
$. In certain sense, the independence of negativity
$E_{13\times24} $ from parameter $ g_{23}t $ represents the
conservation law for nonclassical resources, as the negativities
of the different two-mode reductions derived from this bipartition
($ E_{1\times2} $, $ E_{1\times4} $, $ E_{3\times2} $, and $
E_{3\times4} $) do depend on parameter $ g_{23}t $.

The developed model also allows us to study the role of damping in
the entanglement creation. The investigations based on equal
damping constants $\gamma$ and noiseless reservoirs ($n_{d}=0$)
just reveal the deterioration of entanglement in all the
considered bipartitions with the increase of damping constants
(see Fig.~5).

\section*{Two-mode entanglement and noise reduction factor}

The results of the theoretical analysis suggest that, from
the experimental point of view, the observation of entanglement
between pairs of modes is substantial for the characterization of
the emitted entangled states. Formally, the theory describes such
observations through the reduced two-mode statistical operators.
The analysis shows that the behavior of two-mode negativities $
E_{1\times2} $, $ E_{3\times4} $, and $ E_{1\times4} $ with
respect to parameters $ g_1 t $ and $ g_{23}t $ is qualitatively
similar to that of four-mode negativities $ E_{1\times234} $, $
E_{3\times124} $, and $ E_{12\times34} $ plotted in Figs.~4(a),
(b) and (c). This similarity originates in possible `trajectories'
of photon pairs born in modes 1 and 2 and responsible for the
entanglement.

Additional insight into the generation of entanglement in the
analyzed system is provided when the entanglement is related to
the intensities of the interacting fields. As quantified in the
graphs of Fig.~6, both mean photon numbers $ B_1 \equiv B_2 $ and
$ B_3 \equiv B_4 $ are increasing functions of parameter $ g_1 t $
and oscillating functions of parameter $ g_{23}t $. This
oscillating behavior is particularly interesting, as it reflects
the flow of photons from modes 1 and 2 to modes 3 and 4,
respectively, and vice versa. As we will see below, this is in
agreement with the `flow of the entanglement' among the modes.
\begin{figure} 
\includegraphics[width=0.5\textwidth]{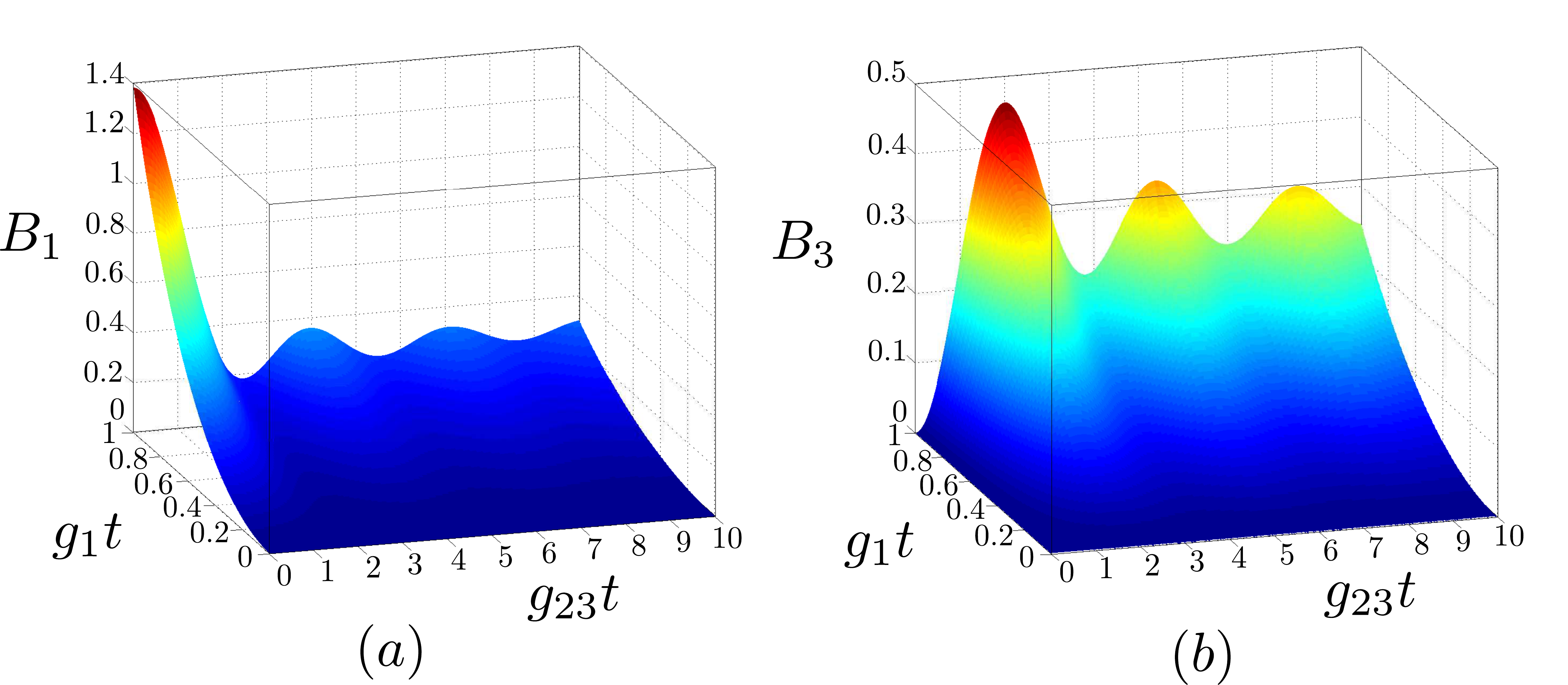}
\caption{Mean photon numbers $B_1$ (a) and $B_3$ (b) plotted as
functions of parameters $g_1t$ and $g_{23}t$.}
\end{figure}

The graph in Fig.~7(a) shows that the negativity $ E_{1\times2} $
is on the one side an increasing function of the mean photon
number $ B_1 $, on the other side it only weakly depends on the
mean photon number $ B_3 $. This confirms that pairing of photons
in parametric down-conversion is the only resource for
entanglement creation. On the contrary, as shown in Fig.~7(b), the
negativity $ E_{3\times4} $ is an increasing function of the mean
photon number $ B_3 $, whereas it weakly depends on the mean
photon number $ B_1 $. This indicates that the entanglement in
modes 34 comes from modes 12 through the transfer of photon pairs:
The stronger the transfer is, the larger the value of negativity $
E_{3\times4} $ is. Moreover, optimal conditions for the
observation of entanglement in modes 1 and 4 occur provided that
there is the largest available number of photon pairs with one
photon in mode 1 and its twin in mode 4. According to the graph in
Fig.~7(c) this occurs when the mean photon numbers $ B_4 $ ($
B_4\equiv B_3 $) and $ B_1 $ are balanced, independently of their
 values.

In general, the experimental identification of two-mode
entanglement is not easy, as it requires the simultaneous
measurement of the entangled state in two complementary bases.
Alternatively, entanglement can be inferred from the reconstructed
two-mode phase-space quasi-distribution, which needs two
simultaneous homodyne detectors \cite{Lvovsky2009}, each one
endowed with a local oscillator. However, the detection of
entanglement, at least in some cases, can be experimentally
accomplished by the observation of sub-shot-noise intensity
correlations. This is a consequence of the detailed numerical
analysis,
%
%
%
%
%
%
\begin{widetext}

\begin{figure}[t!] 
 \includegraphics[width=\textwidth]{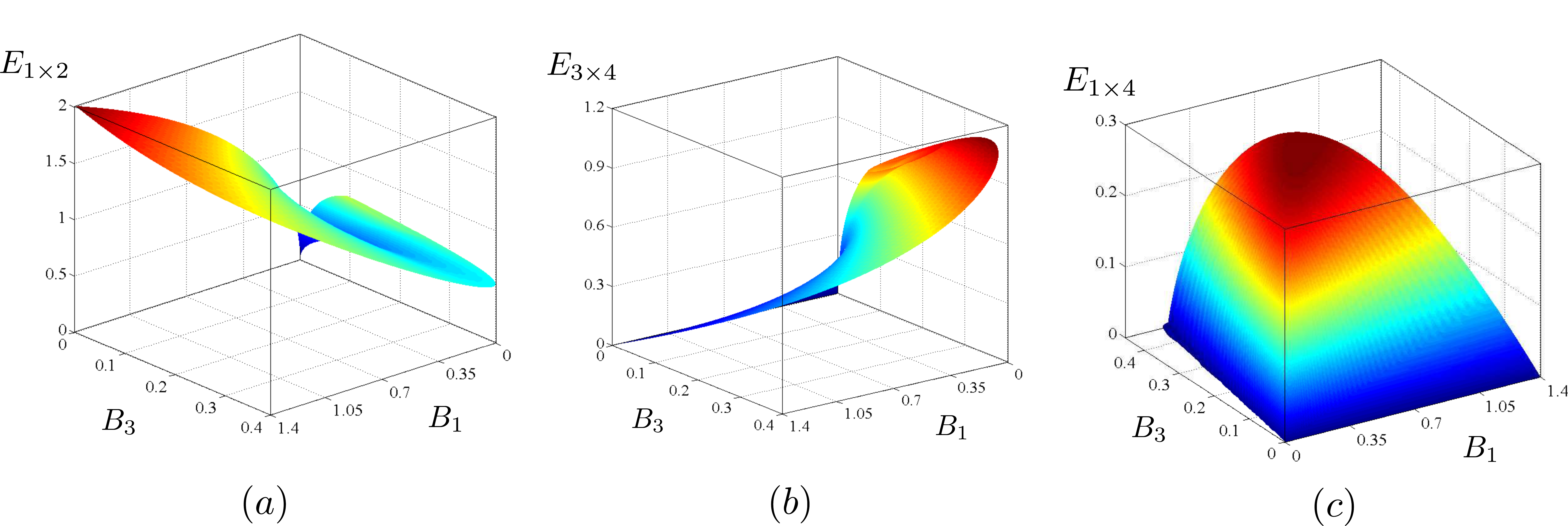}
 \caption{Logarithmic negativities $E_{1\times2} $ (a),
  $E_{3\times4} $ (b) and $E_{1\times4} $ (c) as they depend on mean
  photon numbers $B_1$ and $B_3$.}
  \end{figure}
  \begin{figure}[t!] 
 \includegraphics[width=\textwidth]{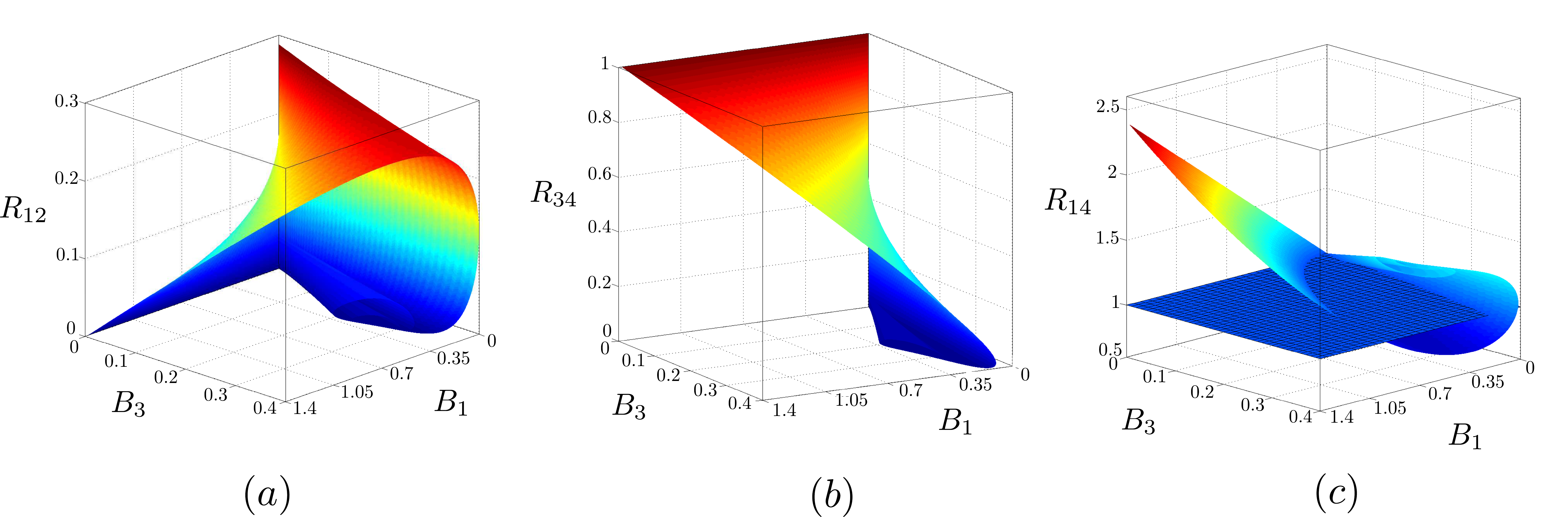}
 \caption{Noise reduction factors $R_{1\times2} $ (a),
  $R_{3\times4} $ (b) and $R_{1\times4} $ (c) as they depend on mean
  photon numbers $B_1$ and $B_3$. In (c), the netted plane is defined
  as $ R_{1\times4}=1 $.}
  \end{figure}
  \end{widetext}
which reveals that the majority of the reduced two-mode entangled
states also exhibits sub-shot-noise intensity correlations.
Nevertheless, it should be emphasized here that, in the analyzed
system, there are also two-mode entangled states not exhibiting
sub-shot-noise intensity correlations. On the contrary, we note
that the reduced two-mode separable states do not naturally
exhibit sub-shot-noise intensity correlations.

Sub-shot-noise intensity correlations are quantified by the noise
reduction factor $R$ \cite{Degiovanni07,Degiovanni09}, that is
routinely measured to recognize nonclassical intensity
correlations of two optical fields. The noise reduction factor $ R
$ expressed in the moments of photon numbers $ n_j $ and $ n_k $
of modes $ j $ and $ k $, respectively, is defined by the formula:
\begin{equation}
 R_{jk} = \frac{\langle\triangle(n_j-n_k)^2\rangle}{\langle
 n_j\rangle+\langle n_k\rangle}.
\label{NRF}
\end{equation}
Sub-shot-noise intensity correlations are described by the
condition $R<1$. We note that there exists the whole hierarchy of
inequalities involving higher-order moments of photon numbers (or
intensities) \cite{Vogel2008,Miranowicz02,Miranowicz10,Allevi2012}
that indicate nonclassicality and, in our system, also
entanglement. We mention here the inequality derived by Lee
\cite{Lee1990a} as a practical example that is sometimes used in
the experimental identification of nonclassicality. We note that
this criterion is stronger than the noise reduction factor $ R $
in revealing the nonclassicality \cite{Degiovanni07}.

The noise reduction factors $R_{1\times2} $, $ R_{3\times4} $ and
$R_{1\times4} $ describing the reduced two-mode fields with their
negativities plotted in Fig.~7 are drawn in Fig.~8 for comparison.
We can see complementary behavior of the negativities $ E $ and
noise reduction factors $ R $ in the graphs in Figs.~7 and 8. An
increase of the negativity $ E $ is accompanied by a decrease in
the noise reduction factor $ R $. A closer inspection of the
curves in these graphs shows that the condition $ R<1 $ identifies
very well entangled states when the noise reduction factor is
measured in modes $ 1\times2 $ and $ 3\times 4 $.
Nevertheless, there are entangled states with $ R_{1\times4}
>1$, as shown in the graph of Fig.~9, in which the values of
parameters $ g_1t$ and $ g_{23}t $ appropriate for this situation
occur in the areas I and III. On the other hand, the entangled
states found in the area II in the graph of Fig.~9 have $ R<1 $.
It is worth noting that the relative amount of entangled states
not detected via $ R < 1 $ increases with the increasing coupling
constant $ g_1 t $ and so with the increasing overall number of
photons in the system.

The observed relation between the entangled states and those
exhibiting sub-shot-noise intensity correlations can even be
explained theoretically, due to the specific form of the reduced
two-mode Gaussian states analyzed in Ref.~\cite{arkhipov15}.
According to Ref.~\cite{arkhipov15} entangled states in modes $ i
$ and $ j $ are identified through the inequality $ B_i B_j <
|D_{ij}|^2 $. On the other hand, the noise reduction factor $
R_{ij} $ defined in Eq.~(\ref{NRF}) attains for our modes the
form:
\begin{equation}
 R_{ij} = 1 + \frac{ B_i^2 + B_j^2 - 2 |D_{ij}|^2 }{ B_i + B_j }
\end{equation}
that assigns the sub-shot-noise intensity correlations to the
states obeying the inequality $ B_i^2 + B_j^2 < 2|D_{ij}|^2 $.
Thus, the inequality $ B_i^2 + B_j^2 \geq 2 B_i B_j $ implies that
the states with sub-shot-noise intensity correlations form a
subset in the set of all entangled states. Moreover, if $ B_i =
B_j $, both sets coincide as we have $ B_i^2 + B_j^2 = 2 B_i B_j
$. Thus, the noise reduction factors $ R_{12} $ and $ R_{34} $ are
reliable in identifying entangled states in the symmetric case, in
which $ B_1 = B_2 $ and  $ B_3 = B_4 $.

We note that, according to the theory developed for the modes
without an additional internal structure \cite{arkhipov15}, the
logarithmic negativity $ E_{ij} $ can be determined along the
formula~\cite{arkhipov15}
\begin{eqnarray}\label{tmE}
 E_{ij}={\rm max} \Big\{0, &&-\log\Big(1+B_i+B_j- \nonumber \\
 &&\sqrt{(B_i-B_j)^2+4|D_{ij}|^2}\Big) \Big\},
\end{eqnarray}
 where $ |D_{ij}|^2 = \langle \Delta
n_i\Delta n_j\rangle $. According to Eq.~\label{tmE} the
logarithmic negativity $ E_{ij} $ can, in principle, be inferred
from the measured mean intensities in modes $ i $ and $ j $ and
the cross-correlation function of intensity fluctuations in this
idealized case.
\begin{figure} 
\includegraphics[width=0.45\textwidth]{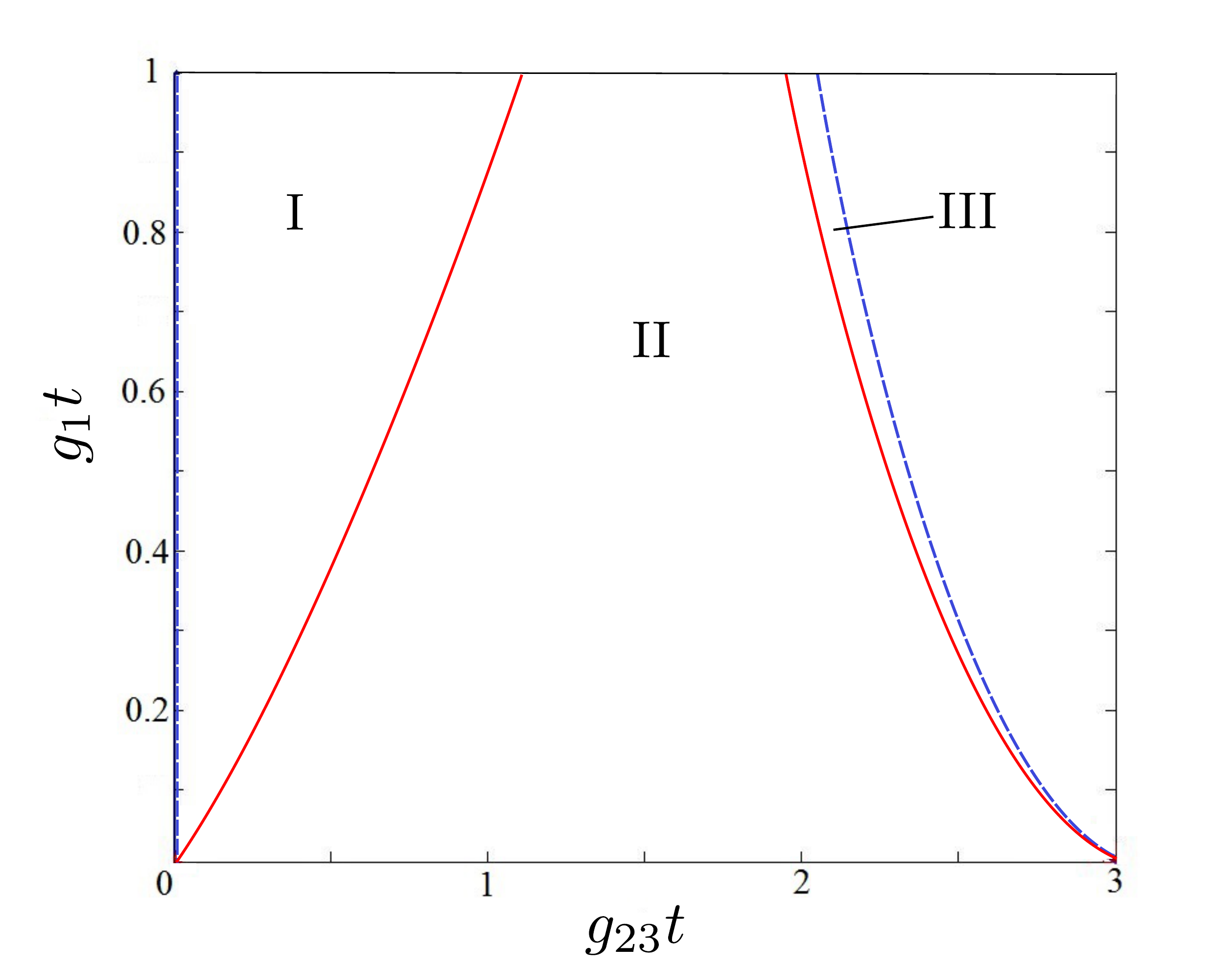}
\caption{Solutions of the equations for logarithmic negativity
$E_{1\times4}=0$ (blue dashed line) and noise reduction factor
$R_{1\times4}=1$ (red solid line) in the plane spanned by
parameters $g_1t$ and $g_{23}t$. The two-mode field is entangled
[sub-shot-noise] ($E_{1\times4}>0$ [$R_{1\times4}<1$]) inbetween
the blue dashed [red solid] lines, i.e. in the areas I, II, and
III [II].}
\end{figure}

At the end, we make a note about the entanglement in the general
four-mode system with different up-conversion coupling constants
($g_2 \neq g_3 $). This is relevant when non-ideal phase-matching
conditions of the three nonlinear interactions are met in the
experiment (see below). According to our investigations, the
largest values of negativities $E_{1\times2}$ and $E_{3\times4}$
are found in the symmetric four-mode system ($ g_2=g_3 $)
considered above. On the contrary, the largest values of
negativities $E_{1\times4}$ and $E_{2\times3}$ are obtained for
unbalanced $ g_2 $ and $ g_3 $ interactions.

Similarly to the symmetric case, separable states, entangled
states without sub-shot-noise intensity correlations and entangled
states exhibiting sub-shot-noise intensity correlations are found
in the whole three-dimensional parametric space spanned by
variables $ g_jt $ for $ j=1,2,3 $. As an example, the
distribution of different kinds of reduced two-mode states found
in the up-converted modes 3 and 4 in this space is plotted in
Fig.~10. The graphs in Fig.~10 indicate that, in accord with the
symmetric case, the larger the value of constant $ g_1 t $, the
larger the relative amount of entangled states that cannot be
identified through sub-shot-noise intensity correlations.
\begin{widetext}

\begin{figure} 
\includegraphics[width=\textwidth]{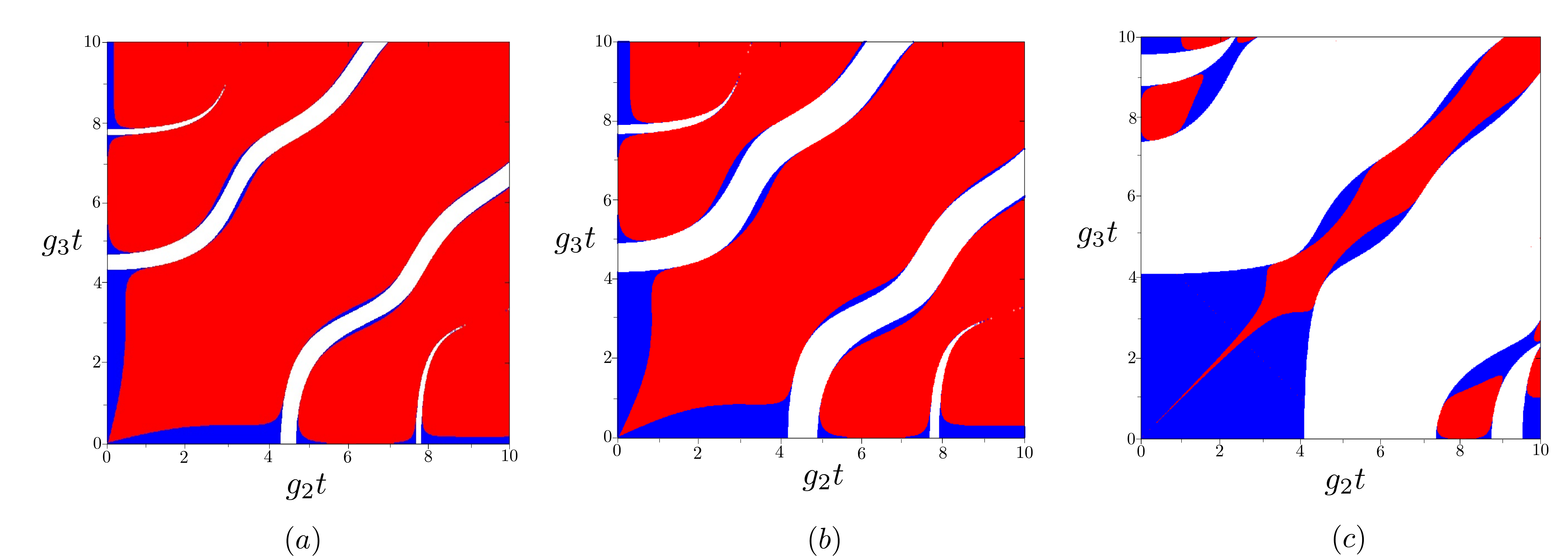}
\caption{ Planes given by $ g_1t = 0.5 $ (a), $
 g_1t = 1 $ (b) and $ g_1t = 5 $ (c) in the 'phase diagram' identifying classical
 states (white areas), entangled states without sub-shot-noise
 intensity correlations (blue) and entangled states with sub-shot-noise
 intensity correlations (red) in the space spanned by the coupling constants
 $ g_j t$, $ j=1,2,3 $.
}
\end{figure}
\end{widetext}

\section*{Experimental implementation}\label{s:exp}

A possible experimental implementation of the four mode
interaction described above can be achieved by using a BaB$ {}_2
$O$ {}_4 $ crystal as the nonlinear medium, a ps-pulsed laser (a
mode-locked Nd:YLF laser regeneratively amplified at 500 Hz,
High-Q Laser Production) to get the pump fields and hybrid
photodetectors (mod. R10467U-40, Hamamatsu Photonics) as the
photon-number-resolving detectors. A typical experimental setup
can be built in analogy with other previous experiments
\cite{Allevi2012}. The phase-matching conditions can be chosen so
as to have $\omega_1=\omega_2$ and a common pump field for both
up-conversion processes so that $\omega_3=\omega_4$. In this
specific symmetric case we have $ g_2 = g_3 \equiv g_{23} $.

We can estimate the range of coupling constants achievable in this
setup based on the above-mentioned laser source. Let us consider
the following parameters: wavelength of the pump for
down-conversion $\lambda_{p1}=349$~nm,
$\lambda_1=\lambda_2=698$~nm, wavelength of the pump for
up-conversion $\lambda_{p2}=1047$~nm,
$\lambda_3=\lambda_4=418.8$~nm, length of the BaB$ {}_2 $O$ {}_4 $
crystal $L=4$~mm, diameters of the pumps 0.5~mm, pulse duration
4.5~ps. The coupling constants $ g_1 $ and $ g_{23} $ are linearly
proportional to the corresponding pump field amplitudes so that $
g_1 t = \kappa_1 A_{p1}L $ and $ g_{23} t = \kappa_{23} A_{p2}L$,
where $\kappa_j$ ($j=1,23$) are the nonlinear coupling
coefficients and $A_j$ ($j=p1,p2$) are the pump amplitudes. For
the considered parameters we can estimate $\kappa_j\approx
10^{-13} s^{1/2}$. The useful range of energies per pulse is up to
66~$\mu$J in the UV and up to 240~$\mu$J in the IR, corresponding
to maximum values $ g_1 t\approx 5.9$ and $ g_2 t\approx 7$. The
theoretical results discussed above predict an interesting
behavior for this range of parameters, including the transfer of
entanglement into the up-converted modes.

\section*{Conclusions}

Four-mode Gaussian states generated via parametric down-conversion
and frequency up-conversion have been analyzed in terms of
nonclassicality, entanglement and entanglement transfer among the
modes. While nonclassicality of the state has been described by
the easily-computable Lee nonclassicality depth, logarithmic
negativity for different bipartitions has been applied to monitor
the occurrence of entanglement among different modes. It has been
shown that whenever the analyzed system is nonclassical, it is
also entangled. Moreover, the entanglement is present only in the
form of bipartite entanglement. The analysis of the noise
reduction factor identifying sub-shot-noise intensity
correlations, in parallel with the logarithmic negativity
quantifying two-mode entanglement, has shown that the noise
reduction factor is a powerful indicator of the entanglement in
the analyzed system. This is substantial for the experimental
demonstration of the transfer of entanglement from the
down-converted modes to the up-converted ones.

 {\bf Acknowledgments} This work was supported by the
projects No.~15-08971S of the GA \v{C}R and No.~LO1305 of the
M\v{S}MT \v{C}R. I.A. thanks project IGA\_2016\_002 UP Olomouc.
The authors also acknowledge support from the bilateral
Czech-Italian project CNR-16-05 between CAS and CNR.

 {\bf Author contributions statement} I.A., J.P., O.H.,
A.A. and M.B. developed the theory. I.A. prepared the figures.
I.A., J.P., O.H., A.A. and M.B. wrote the manuscript. All authors
reviewed the manuscript.

\newpage
\appendix
\appendix
\begin{widetext}
\section{The evolution matrix \protect{ $ {\bf M} $}}

The evolution matrix $ {\bf M} $ describing the operator solution
of the Heisenberg equations written in Eq.~(\ref{hle}) is derived
in the form:
\begin{eqnarray}\label{M}
 {\bf M} = \Large \left( \begin {array}{cccc} {\frac {x{ c_1}-y{c_2}}{x-y} }&{\frac
 {ixy \left( \sqrt {{ y_1}}{s_2}-\sqrt {{x_1}}{
  s_1} \right) }{ \left( x{ y_1}-{ x_1}\,y \right) { g_1}}}
 &{\frac {i \left( y\sqrt {{y_1}}{ x_1}\,{ s_2}-x\sqrt {{
  x_1}}{y_1}\,{s_1} \right) }{{ g_2}\, \left( x{y_1}-{
  x_1}\,y \right) }}&{\frac {xy \left( {c_2}-{c_1}
  \right) }{{g_1}\,{g_3}\, \left( x-y \right) }}
 \\ \noalign{\medskip}{\frac {i{g_1}\, \left( \sqrt {{y_1}
 }{s_2}-\sqrt {{ x_1}}{s_1} \right) }{x-y}}&{\frac {x{ y_1}\,{
 c_2}-{x_1}\,y{c_1}}{x{y_1}-{x_1}\,y}} &{\frac
 {{g_1}\,{y_1}\,{x_1}\, \left( {c_2}-{ c_1} \right) }{{g_2}\,
 \left( x{y_1}-{x_1}\,y \right) }}&{ \frac { i \left( \sqrt
 {{y_1}}x{ s_2}-\sqrt {{x_1}}y{ s_1} \right) }{ \left( x-y \right)
 {g_3}}}
 \\ \noalign{\medskip}{\frac {i{ g_2}\, \left( \sqrt {{ x_1}}y{
 s_2}-\sqrt {{y_1}}x{s_1} \right) }{\sqrt {{x_1}}
  \left( x-y \right) \sqrt {{y_1}}}}&-{\frac {{g_2}\,xy \left( {
 \it c_2}-{c_1} \right) }{ \left( x{y_1}-{x_1}\,y
  \right) {g_1}}}&{\frac {x{y_1}\,{c_1}-{x_1}\,y{
 c_2}}{x{ y_1}-{x_1}\,y}}&{\frac {-i{ g_2}\,xy \left( \sqrt {
 {x_1}}{s_2}-\sqrt {{y_1}}{s_1} \right) }{{ g_1} \,{g_3}\, \left(
 x-y \right) \sqrt {{ x_1}}\sqrt {{y_1}}}}
 \\ \noalign{\medskip}{\frac {{g_1}\,{g_3}\, \left( -{c_2}
 +{c_1} \right) }{x-y}}&{\frac {i{ g_3}\, \left( \sqrt {{ y_1}}x{
 s_2}-\sqrt {{ x_1}}y{ s_1} \right) }{x{y_1}-{
  x_1}\,y}}&{\frac {i{ g_1}\,{g_3}\, \left( { x_1}\,\sqrt {{
  y_1}}{s_2}-{ y_1}\,\sqrt {{ x_1}}{ s_1} \right) }{
 {g_2}\, \left( x{ y_1}-{x_1}\,y \right) }}&{\frac {x{
 c_2}-y{ c_1}}{x-y}}\end {array} \right) \nonumber \\
\end{eqnarray}
where $x=(a+b)/2$, $x_1=(a_1+b)/2$, $y=(a-b)/2$, $y_1=(a_1-b)/2$,
$a=-g_1^2+g_2^2-g_3^2$, $a_1=-g_1^2+g_2^2+g_3^2$,
 $b=\sqrt{{{ g_1}}^{4}-2\,{{ g_1}}^{2}{{ g_2}}^{2}-2\,{{g_1}}^{2}{{g_3}}^{2}+{{g_2}}^{4}-2\,{{ g_2}}^{2}{{g_3}}^{2}+{{g_3}}^{4}}$,
 $c_1 = \cos(\sqrt{x_1}t) $, $c_2 = \cos(\sqrt{y_1}t) $, $s_1 = \sin(\sqrt{x_1}t) $, and $ s_2 = \sin(\sqrt{y_1}t) $.

\end{widetext}


\end{document}